\documentclass[preprint2]{aastex61}
\usepackage{float}
\usepackage{multirow}
\shorttitle{Ocean Circulations on Exoplanets}
\shortauthors{Ji, Chen and Yang}

\begin{document} 

\title{Idealized Wind-driven Ocean Circulations On Exoplanets}

\correspondingauthor{Jun Yang}
\email{junyang@pku.edu.cn}

\author[0000-0002-5366-6880]{Weiwen Ji}
\affiliation{Department of Atmospheric and Oceanic Sciences, School of Physics, Peking University, 100871, Beijing, China}

\author{Ru Chen}
\affiliation{University of California, 92521, Los Angeles, USA}

\author{Jun Yang}
\altaffiliation{}
\affiliation{Department of Atmospheric and Oceanic Sciences, School of Physics, Peking University, 100871, Beijing, China}





\
\
\begin{abstract}

Motivated by the important role of the ocean in the Earth climate system, here we investigate possible scenarios of ocean circulations on exoplanets using a one-layer shallow water ocean model. Specifically, we investigate how planetary rotation rate, wind stress, fluid eddy viscosity and land structure (a closed basin vs. a reentrant channel) influence the pattern and strength of wind-driven ocean circulations. The meridional variation of the Coriolis force, arising from planetary rotation and the spheric shape of the planets, induces the western intensification of ocean circulations. Our simulations confirm that in a closed basin, changes of other factors contribute to only enhancing or weakening the ocean circulations (e.g., as wind stress decreases or fluid eddy viscosity increases, the ocean circulations weaken, and vice versa). In a reentrant channel, just as the Southern Ocean region on the Earth, the ocean pattern is characterized by zonal flows. In the quasi-linear case, the sensitivity of ocean circulations characteristics to these parameters is also interpreted using simple analytical models. This study is the preliminary step for exploring the possible ocean circulations on exoplanets, future work with multi-layer ocean models and fully coupled ocean-atmosphere models are required for studying exoplanetary climates.


\end{abstract}




\section{Introduction} \label{sec:intro}

Recently, much attention has been paid to exoplanets, and ``the trickle of discoveries has become a torrent'' \citep{hecht2016truth}. An exoplanet, also termed as extrasolar planet, is a planet beyond our solar system that orbits a star. According to the latest data from NASA's Exoplanet Archive, over 3700 exoplanets have been confirmed (\url{https://exoplanets.nasa.gov/}). The ultimate goal of exoplanet detection is to find other habitable planets outside the solar system and even other intelligent lives in the universe. Among numerous exoplanets, \citet{leger2004new} suggested the presence of ocean planets around other stars. In 2006, the discovery of the cool planet OGLE-2005-BLG-390Lb \citep{beaulieu2006discovery} indicated the opportunity to detect ocean planets in the future missions. These ocean planets are supposed to have very deep oceans (could be as deep as 100 $km$), hence a lower value of planetary density compared to rocky planets.

Though the oceans on exoplanets remain largely unknown, the Earth ocean has been extensively studied and is found to be key to the Earth climate. The Earth ocean stores and exchanges tremendous amount of water, heat and biogeochemical tracer with the atmosphere and cryosphere. In particular, ocean circulations have large impacts on climate by transporting energy poleward from the tropics to the poles. \citet{trenberth2001estimates} have shown that, though small at mid- and high-latitudes, the magnitude of oceanic heat transport in the tropics is comparable to that of the atmospheric heat transport. When the ocean circulations change (i.e., weaken or strengthen), sea surface temperature and sea ice distribution would have corresponding responses. For example, sea ice would extend to larger areas if ocean heat transport is turned off, and thereby surface temperatures at high latitudes would drop significantly \citep{winton2003climatic}.

Considering the key role of ocean circulations in the Earth climate, it is reasonable to expect that ocean circulations would be important to exoplanetary climates as well, if ocean exists there. For example, ocean circulations have been proved to be crucial in determining the climate and habitability of exoplanets \citep{hu2014role,cullum2016importance,del2017habitable}.

Our goal is to evaluate the ocean circulations on exoplanets, based on our knowledge of the ocean circulations on the Earth. Ocean circulations on the Earth have two components: the wind-driven circulations and deep thermohaline circulations. The former is energetic and mainly locates in the upper ocean, whereas the latter is relatively sluggish and can reach the deep ocean \citep{huang2010ocean}. \citet{boccaletti2005vertical} demonstrated that the wind-driven circulations dominate the meridional oceanic heat transport. Specifically, They found that the shallow circulations in the upper 500 meters driven by wind stresses contribute to nearly all of the heat transport in the Southern Hemisphere, and the wind-driven circulations also predominate the heat transport in the Northern Hemisphere. Therefore, we focus on the wind-driven circulations only in this paper.

Two typical types of continental configuration used to study idealized wind-driven ocean circulations on the Earth are closed basin and channel model. In the closed ocean basin, a system of circulating currents will form, which is termed an ocean gyre. Western intensification, which refers to an intense and narrow western boundary current (e.g., the Gulf Stream in the North Atlantic and the Kuroshio in the North Pacific Ocean, shown in Figure 1(a)) is the most remarkable feature of an ocean gyre. The origin of the western intensification is related to the meridional variation of the Coriolis parameter \citep{stommel1948westward,munk1950wind}. Among all the ocean currents, the western boundary current contributes the most to the poleward heat transport and plays a significant role in maintaining the global heat balance \citep{dunxin1991western}. In the channel model, a zonal current can arise because of the zonal wind forcing and this model has been used extensively to study the Antarctic Circumpolar Current (ACC) in the Southern Ocean (e.g., \citealt{nadeau2015role}). The ACC, as a major feature of the circulations in the Southern Ocean, circulates around the Antarctica and is very energetic (see Figure 1(a)). It is the strongest current in the oceans \citep{pickard2016descriptive} and has a zonal transport of 98-154 \emph{Sv} (1 \emph{Sv} = 10$^{6}$ \emph{m}$^{3}$ \emph{s}$^{-1}$, \citealt{whitworth1985volume}), larger than the transport of the Gulf Stream (approximately 31 \emph{Sv} of upper-layer wind-driven part, \citealt{lund2006gulf}).

Though fundamental fluid dynamics probably hold for both the Earth and exoplanet ocean, the ocean circulations structure on exoplanets can be quite different from those on the Earth. Our goal is to assess possible scenarios of the wind-driven circulations on exoplanets and compare them with those on the Earth. Specifically, we use a one-layer shallow water model to evaluate how exoplanet-relevant model parameters can lead to different spatial structure and strength of wind-driven circulations in both a closed basin and a channel. The parameters we consider include planetary $\beta$ (approximate variation of the Coriolis parameter with latitude), wind stress, viscosity and ocean basin structure (closed or open in the zonal direction). Our work is built on previous knowledge of the wind-driven circulations on the Earth (e.g., \citealt{charney1955gulf,lund2006gulf,hu2015pacific,mcwilliams1978description,orsi1995meridional}).

This paper is organized as follows. Sections 2 and 3 introduce the numerical model and describe the setup of the numerical experiments. Results and interpretations are presented in Sections 4 and 5. Section 6 provides the summary and discussion.

\
\section{Model descriptions} \label{sec:style}

We use a one-layer shallow water model to simulate the barotropic wind-driven ocean circulations. The shallow water approximation is used when the horizontal scale of the fluid is much larger than its depth, which implies that the large-scale vertical velocities are much smaller than horizontal velocities. 
The shallow water model is one of the most useful models in geophysical fluid dynamics, especially for studying the ocean and atmosphere \citep{vallis2017atmospheric}. The one-layer shallow water model, which is one of the simplest model to produce ocean circulations, can reveal fundamental ocean dynamics and at the same time allows ease of interpretation. Recent work indicates that idealized shallow water model is useful for exploring fluid structures on giant planets and successfully predicts the existence of polar cyclones on Jupiter \citep{o2015polar,o2016weak}.

This model is from one component set of the MIT General Circulation Model (MITgcm, \citealt{ferreira2006formulation, marshall2007mean}). The ocean circulations are driven by surface wind stresses and dissipated by eddy viscosity. The model is configured to represent a square enclosed box of water with a horizontal length \emph{L} of 1,200 \emph{km} $\times$ 1,200 \emph{km}, a vertical depth of 5 \emph{km} and a horizontal resolution of 20 \emph{km}. Lateral eddy viscous dissipation is included in the model. The governing equations are

\begin{eqnarray}
\frac{Du}{Dt}-fv+g\frac{\partial \eta}{\partial x}-A_{h}\bigtriangledown_{h}^{2}u &=& \frac{\tau_{x}}{\rho_{0}\bigtriangleup{z}},   \\
\frac{Dv}{Dt}+fu+g\frac{\partial \eta}{\partial y}-A_{h}\bigtriangledown_{h}^{2}v &=& 0,      \\
\frac{\partial \eta}{\partial t}+\frac{\partial uH}{\partial x}+\frac{\partial vH}{\partial y} &=& 0,
\end{eqnarray}
where \emph{u}, \emph{v} are the \emph{x} and \emph{y} zonal and meridional velocities; $\frac{D}{Dt}=\frac{\partial}{\partial t}+u\frac{\partial}{\partial x}+v\frac{\partial}{\partial y}$ is the horizontal material derivative in the Cartesian coordinate; $f=f_{0}+\beta{y}$ is the Coriolis parameter, and $\beta=\frac{\partial{f}}{\partial{y}}=\frac{2{\Omega}\cos{\theta_0}}{R}$ is the meridional variation of the Coriolis parameter at the latitude $\theta_0$ ($\Omega$ is the planetary rotation rate and \emph{R} is the planetary radius); $\eta$ represents sea surface height;  $g=9.81$ $m$ $s^{-2}$ is the acceleration due to gravity; $A_h$ is the eddy viscosity coefficient; $\bigtriangledown_{h}^{2}=\frac{\partial^{2}}{\partial x^2}+\frac{\partial^2}{\partial y^2}$ is the horizontal Laplacian operator; $\rho_0$ is the reference water density, and we assume the density is constant; $\bigtriangleup{z}=5$ $km$ is the mean ocean depth; $\tau_x$ is the zonal wind stress; meridional wind is neglected in our experiments; and $H=\bigtriangleup{z}+\eta$ is the entire ocean depth.

\
\section{Experimental design} \label{sec:style}
Similar to the studies of Earth ocean circulations, we set the experiments into two types of terrains: a closed basin and a channel ocean. We examine the effects of three parameters, the variation of the Coriolis parameter ($\beta$), the viscosity parameter ($A_h$) and the surface wind stress ($\tau$), on the ocean circulations. The choice of parameters is motivated by the fact or possibilities that these exoplanets have sizes, rotation rates and thus $\beta$ different from the Earth, and their ocean fluid viscosity, atmospheric circulations and thus wind stress can also be different from those of the Earth.

In the closed basin, the ocean is surrounded by land at all four boundaries, whereas in the channel model, the land locates in the northern/southern boundary only. We use a no-slip and no normal flow boundary condition at the four boundaries of the closed basin and at the northern/southern boundary of the channel, that is, the velocities there are set to be zero. In the channel model, a periodic boundary condition is employed in the zonal direction, that is, the ocean is reentrant. The ranges of the parameters ($\beta$, $\tau$, and $A_h$) we investigate are 0.1, 0.5, 2, or 10 times the default values (see Table 1).

We use zonal wind stress only, with a form of $\tau_x=-\tau \cos{\frac{\pi y}{L}}$ (see Figure 1(b)), which is similar to the wind stress in the subtropics of the Earth. All the experiments reach steady state after forty years, except for the cases with the onset of instability.

\begin{table}[h]
\caption{Experimental Arrangements}
\renewcommand\arraystretch{1.2} 
\centering
\begin{tabular}{p{1.8cm}p{1.6cm}<{\centering}p{0.6cm}<{\centering}p{4.7cm}<{\centering}}
\hline
Oceans & Parameters & Runs & Design \\
\hline
\multirow{6}{*}{Closed basin} & \multirow{3}{*}{Control} & \multirow{3}{*}{1} & planetary $\beta$: $10^{-11}$ $m^{-1}$ $s^{-1}$   \\
  & & & viscosity $A_h$: 400 $m^{2}$ $s^{-1}$ \\
  & & &  wind stress $\tau$: 0.1 $N$ $m^{-2}$    \\
  &  $\beta$    &    3     &    2$\textrm{,}$ $\frac{1}{2}$$\textrm{,}$ $\frac{1}{10}$ $\times$ $10^{-11}$ $m^{-1}$ $s^{-1}$   \\
  &  $A_h$      &    3     &    2$\textrm{,}$ $\frac{1}{2}$$\textrm{,}$ $\frac{1}{10}$ $\times$ $400$ $m^{2}$ $s^{-1}$    \\
  &  ${\tau}$     &    3     &    2$\textrm{,}$ $\frac{1}{2}$$\textrm{,}$ 10 $\times$ $0.1$ $N$ $m^{-2}$ \\
\hline
\multirow{8}{*}{Channel ocean} & \multirow{4}{*}{Control} & \multirow{4}{*}{1} & no east-west boundaries  \\
  & & &  planetary $\beta$: $10^{-11}$ $m^{-1}$ $s^{-1}$    \\
  & & & viscosity $A_h$: 400 $m^{2}$ $s^{-1}$ \\
  & & &  wind stress $\tau$: 0.1 $N$ $m^{-2}$    \\
  &  $\beta$    &    2     &    2$\textrm{,}$ $\frac{1}{2}$ $\times$ $10^{-11}$ $m^{-1}$ $s^{-1}$   \\
  &  $A_h$      &    2     &     2$\textrm{,}$ $\frac{1}{2}$ $\times$ $400$ $m^{2}$ $s^{-1}$    \\
  &  ${\tau}$     &    2     &   2$\textrm{,}$ $\frac{1}{2}$ $\times$ $0.1$ $N$ $m^{-2}$ \\
  &    Turbulence &    1     &    barotropic instability onset \\
\hline
\end{tabular}
\end{table}

\begin{figure}[h]
\centering
\includegraphics[width=3.4in]{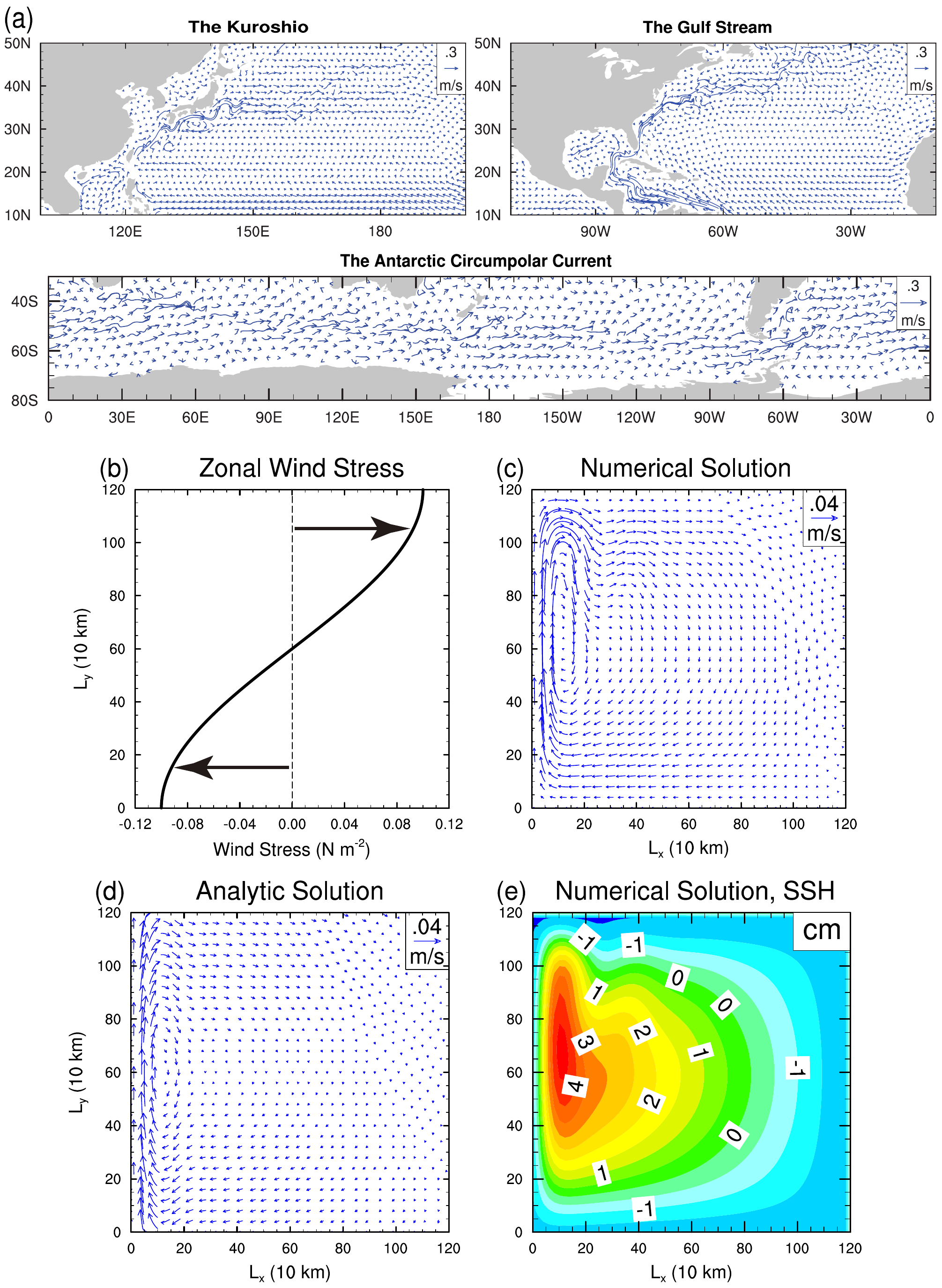}
\caption{Ocean flows in observation of Earth and the control simulation of the model. (a) Observed ocean surface currents (5 \emph{m} below sea level) in three regions, the Kuroshio, the Gulf Stream and the Antarctic Circumpolar Current (annual-mean data of 2009; \citealt{carton2018updated}). (b) The zonal wind stress specified in the model. (c) Simulated steady flows. (d) Analytic solution when the advection terms are neglected \citep{vallis2017atmospheric}. (e) Simulated results of sea surface height (SSH).}
\end{figure}

\
\section{Results of a closed basin} \label{sec:style}

Figure 1(c) shows the flow in a closed basin from the control experiment, driven by the wind stress with a cosine profile (Figure 1(b)). It is characterized by a clockwise gyre: In the oceanic interior, the flow is southward due to the Sverdrup balance, which denotes the vorticity balance between the meridional advection of planetary vorticity and wind stress curl \citep{vallis2017atmospheric}, roughly holds in the oceanic interior from both the control experiment here and the mid-latitude ocean \citep{wunsch2011decadal}. Therefore, the southward flow in the oceanic interior is due to vorticity input from the negative wind stress curl.

Note that the gyre flow is asymmetric in the zonal direction: the oceanic flow at the western boundary is much stronger than that at the ocean interior and the eastern boundary. This phenomenon is termed as ``western intensification'' \citep{gill1982atmosphere} and it is due to the meridional variation of the Coriolis parameter ($\beta$), induced by rotation and the spherical shape of the Earth \citep{stommel1948westward}. Similar to the velocity vectors, the sea surface height also has a gyre structure with the maximum value and the largest slope near the western boundary (see Figure 1(e)). The consistency between the sea surface height and velocity patterns is because of geostrophic balance, which denotes the balance between the Coriolis force and pressure gradient force and roughly holds for large-scale flows. The result of the experiment is generally consistent with the analytic solution shown in Figure 1(d), when the nonlinear advection terms are small.

Both the magnitude and meridional structure of the wind stress we used in our control experiment (Figure 1(b)) are similar to those in the subtropical gyre region in the Northern Hemisphere of the Earth Ocean. The simulated ocean currents, however, are about one order of magnitude weaker than the observations (comparing Figures 1(c) with 1(a)). The reason is that the ocean depth is set to 5 \emph{km} in our experiments, which is much greater than the depth of wind-driven ocean circulations on Earth, generally less than 1 \emph{km} \citep{talley2011descriptive}. Consistently, the simulated sea surface heights (Figure 1(e)) are one order smaller than the observations. Note that, though the strength of the gyre flow depends on  the choice of ocean depth, the sensitivity of the oceanic circulations to parameters, presented next, is not sensitive to the choice of ocean depth.

\begin{figure}[h]
\centering
\includegraphics[width=3.4in]{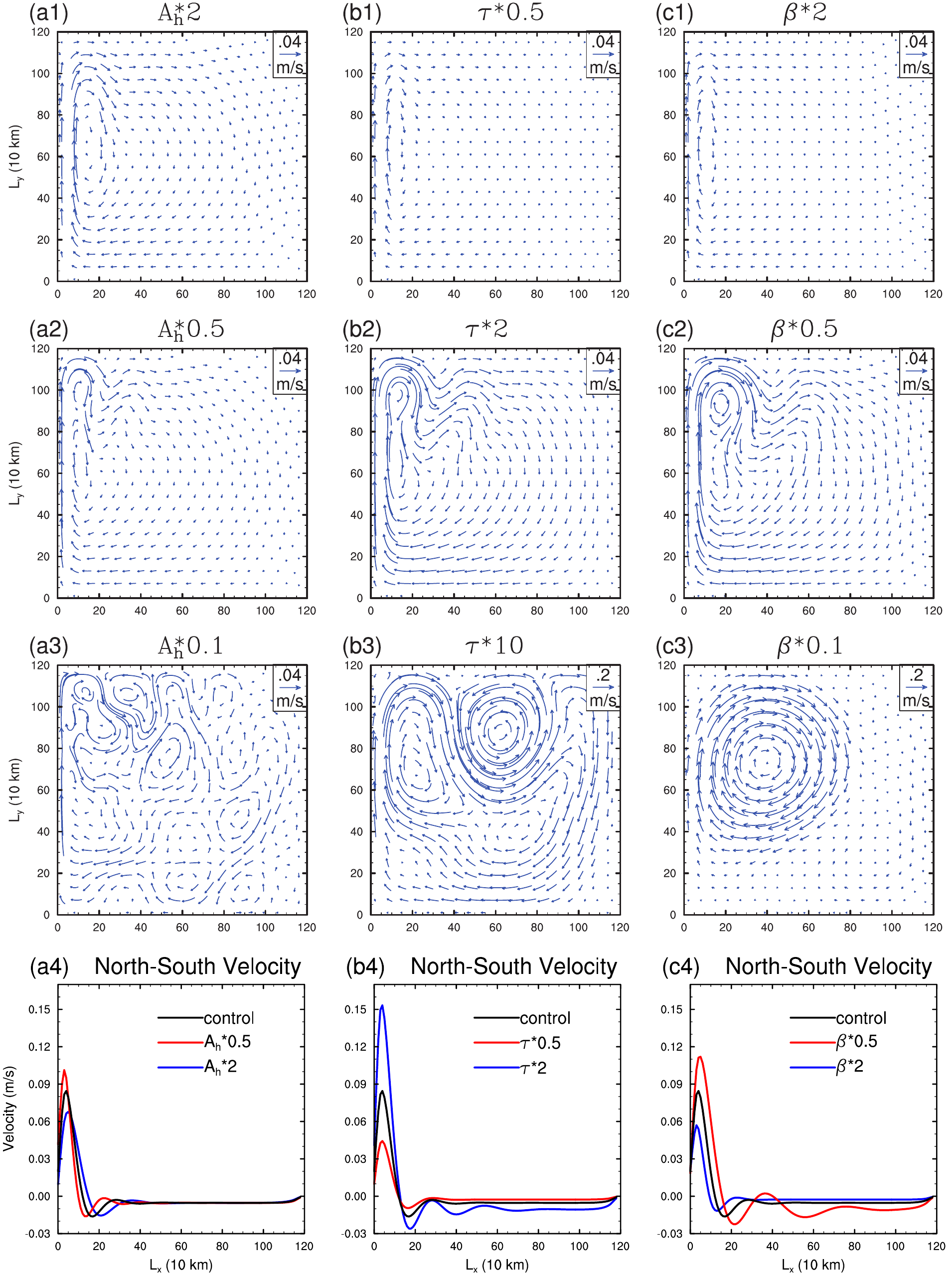}
\caption{Influences of different parameters on the flow field in a closed basin. (a1-a4) Varying the viscosity to 2, 0.5, and 0.1 of the default value. (b1-b4) Varying the wind stress to 0.5, 2, and 10 of the default value. (c1-c4) Varying the $\beta$ to 2, 0.5, and 0.1 of the default value. (a3, b3, and c3) Snapshots of unsteady states. (a4, b4, and c4) Profiles of north-south (meridional) velocities in the middle ($y=600$ \emph{km}) of the ocean.}
\end{figure}

Figure 2 shows the sensitivity of the oceanic circulations to three model parameters: $\beta$, viscosity coefficient $A_h$ and the magnitude of the wind stress $\tau$. The western boundary layer thickness, velocity magnitude and circulation patterns do change with those parameters. First, as $A_h$ increases or $\beta$ decreases, the width of the western boundary layer gets larger, and vice versa (Figures 2(a4) and 2(c4)). However, the western boundary layer thickness is not sensitive to the wind stress magnitude (Figure 2(b4)). Second, consistent with the momentum balance, when decreasing $A_h$ or increasing $\tau$, the ocean currents become stronger, and vice versa (Figures 2(a1-a3) and 2(b1-b3)). We also found that, as $\beta$ increases (decreases), the current speed decreases (increases) (Figures 2(c1)-(c4)).

The sensitivity of the oceanic circulations to those parameters, described above, can be interpreted from the vorticity budget under the quasi-geostrophic assumption \citep{vallis2017atmospheric}, which roughly holds here. First, assuming a balance between friction and meridional advection of planetary vorticity, the characteristic width of the western boundary layer thickness is $L_b \sim {(\frac{A_h}{\beta})}^{\frac{1}{3}}$ \citep{munk1950wind}. The value of $L_b$ in the control experiment is 34 \emph{km}, which is roughly consistent with the numerical result. As $A_h$ increases or $\beta$ decreases, $L_b$ increases; consistently, the western boundary layer thickness from the numerical experiment also increases (Figures 2(a4) and 2(c4)). Second, in the oceanic interior, the Sverdrup balance holds in the linear case, that is ${\beta}v=curl(\frac{{\tau}_x}{\rho_{0}})$, where \emph{v} is the meridional velocity and $curl({\tau}_x)$ is the wind stress curl. Thus, as $\beta$ decreases or wind stress gets larger, the southward oceanic current in the oceanic interior gets stronger, and consistently to conserve mass, the northward western boundary current gets stronger.

When $A_h$ and $\beta$ are larger or the wind stress magnitude is smaller, the equilibrated oceanic flow field is steady, with a gyre structure similar to that in the control experiment (Figures 2(a1), (b1) and (c1)).On the other hand, if $A_h$ and $\beta$ are smaller or the wind stress magnitude is larger, the equilibrated oceanic flow is turbulent with eddies (Figures 2(a3), (b3) and (c3)). Strong nonlinearity leads to turbulence and from a quasi-geostrophic vorticity balance perspective, the degree of nonlinearity can be qualified by $\frac{U}{{\beta}L^2}$, where \emph{U} and \emph{L} are characteristic velocity and length scales \citep{vallis2017atmospheric}. Smaller $\beta$ and $A_h$ and larger wind stress magnitude correspond to stronger nonlinearity, leading to turbulence flow.

\begin{figure}[h]
\centering
\includegraphics[width=3.4in]{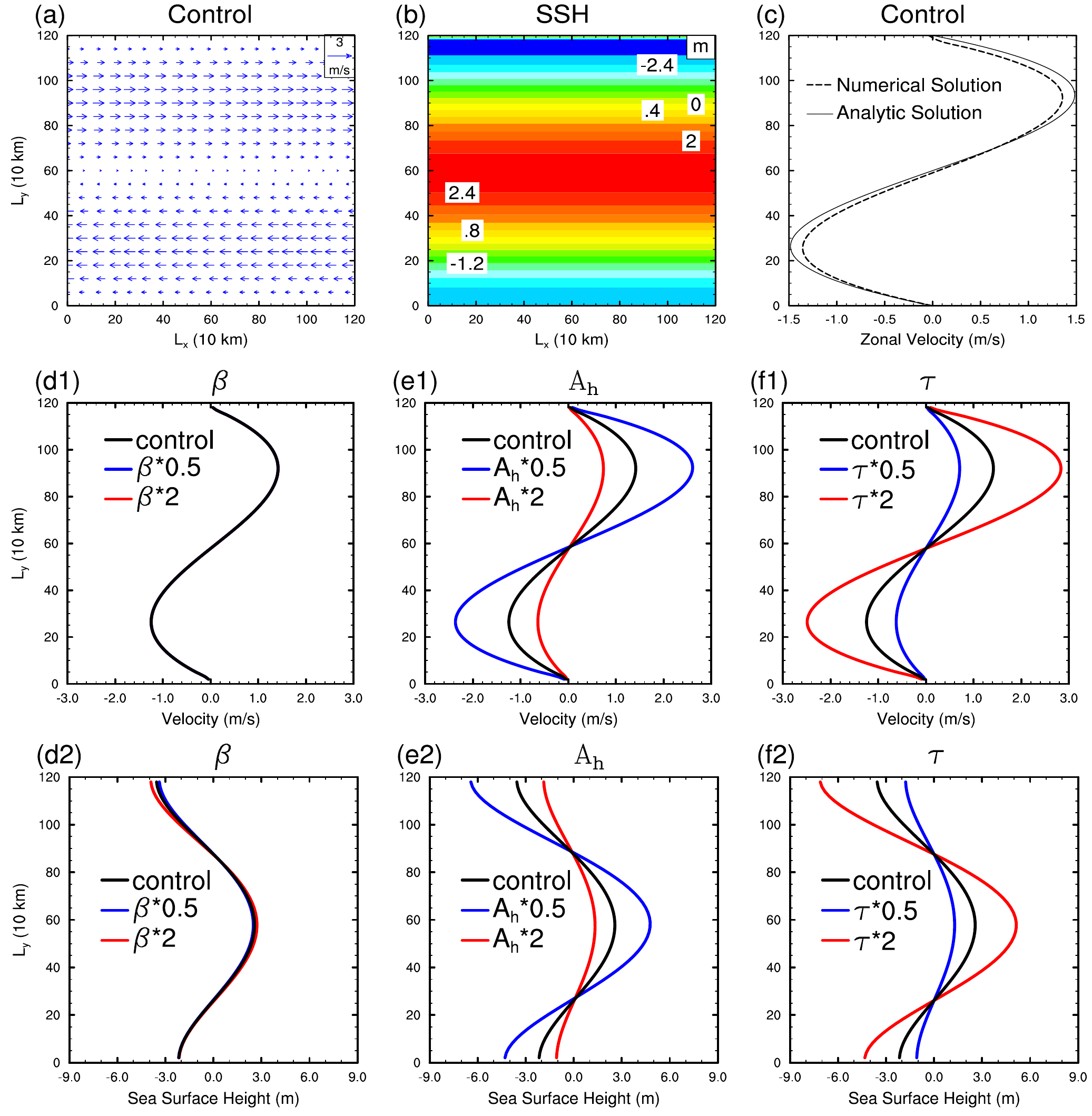}
\caption{Ocean flows in a channel ocean. (a) The ocean velocities in the control run. (b) Sea surface height in the control run. (c) Comparisons between the numerical result and the analytic solution. Zonal-mean ocean currents and sea surface height when varying $\beta$ (d1-d2), varying the viscosity (e1-e2), and varying the wind stress (f1-f2).}
\end{figure}

\section{Results of a channel ocean} \label{sec:style}

Figures 3(a) and 3(b) show the oceanic flow and sea surface height in a channel, forced by the wind stress same as that used in the closed-basin control experiment. Here the oceanic circulations are dominated by zonal flow, which is generally much stronger than those in the closed basin mainly due to the absence of meridional boundaries. The zonal flow is westward in the southern part of the domain and eastward in the northern part. Its minimum value occurs at the northern and southern boundaries, due to no-slip condition, and at the center of the channel, where the wind stress is weak. Sea surface height reaches maximum at the middle of the channel, which is consistent with the geostrophic balance in the meridional direction.

The oceanic circulation in this experiment can be well predicted using a simple analytical model. In the steady state, assuming there is neither meridional velocity nor zonal variation, the governing equations of the model (Equations (1) and (2)) can be reduced to,

\begin{eqnarray}
A_h\frac{\partial^2 u}{\partial y^2} &=& \frac{\tau_{x}}{\rho_{0}\bigtriangleup{z}} ,  \\
fu+g\frac{\partial \eta}{\partial y} &=& 0.
\end{eqnarray}
Then we can obtain an analytical solution by solving Equation (4) with the no-slip boundary condition ($u_{y=0}=u_{y=L}=0$) at the northern/southern boundary,

\begin{equation}
u=\frac{-L^{2}\tau}{\pi^{2}A_h\rho_{0}\bigtriangleup{z}}(\cos{(\frac{\pi{y}}{L})}+\frac{y}{L}-1).
\end{equation}
This analytical solution agrees well with the simulated zonal velocity (Figure 3(c)), and the slight mismatch could be attributed to numerical dissipations in the model. Equation (6) reveals that the zonal flow is not sensitive to the choice of $\beta$, and it increases with the decrease of $A_h$ or increase of $\tau$. These are both confirmed by our numerical experiments (Figures 3(d1), (e1) and (f1)).

Figures 3(d2), 3(e2) and 3(f2) show the meridional profile of sea surface height. In the meridional direction, the momentum equation is reduced to the geostrophic balance (Equation (5)), and thus, the meridional slope of sea surface height increases as the zonal velocity magnitude increases. Therefore, both zonal velocity and sea surface height slope increase, as $A_h$ decreases or $\tau$ increases. Note that although the zonal velocity is insensitive to the choice of $\beta$, the sea surface height profile slightly depends on $\beta$. This is because the Coriolis parameter \emph{f} is equal to $f+{\beta}y$, and the magnitude of $\beta{y}$, $\sim 10^{-5}$ $s^{-1}$, is one order smaller than the first part $f_0$, $\sim 10^{-4}$ $s^{-1}$.

When we accelerate the flows in the channel ocean by reducing $A_h$ or increasing $\tau$, the flow field remains to be zonal currents without turning into an unsteady state. Instability occurs when we change the form of wind stresses (Figure 4(a1)), which drives the zonal currents to meet the necessary condition for barotrpic instability: the expression $\beta-\frac{\partial^{2}u}{\partial y^2}$ changes its sign (Figure 4(a2), \citealt{rayleigh1879stability}). In this experiment, the adjusted wind stress is concentrated in the middle part of the ocean and its magnitude enlarges to 0.5 $N$ $m^{-2}$, five times larger than that in control experiment. Meridional random fluctuations with a magnitude of 0.01 $N$ $m^{-2}$ are added to trigger the instability. When the model runs for 160 days, the instability occurs (Figure 4(a3)). After 400 model days, the flow becomes large-scale waves (Figures 4(a4-a6)). The wavelength developed by the barotropic instability in the model may be limited by the model domain size in the east-west direction.

\begin{figure}[h]
\centering
\includegraphics[width=2.45in,angle=90]{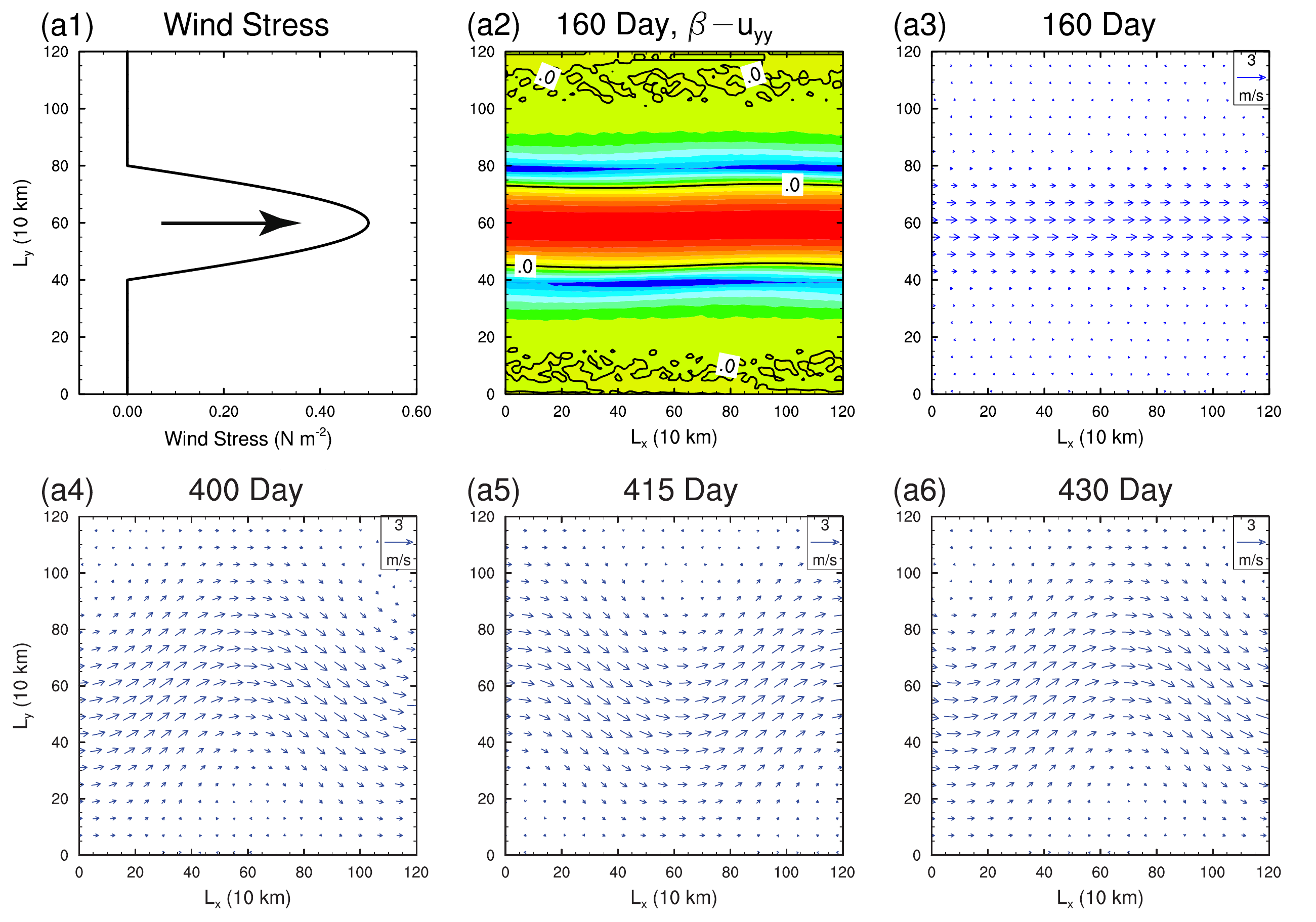}
\caption{Specified zonal wind stress and the development of barotropic instability in a channel ocean. (a1) Wind stress $\tau_x=\tau \sin{\frac{\pi (3y-L)}{L}}\textrm{,}$ $y\in [\frac{L}{3}\textrm{,}\frac{2L}{3}]$. (a2) A snapshot of $\beta{-}u_{yy}$ on $160^{th}$ day, where warm colors mean positive values and cool colors mean negative values. (a3) Snapshots of ocean currents on $160^{th}$, $400^{th}$, $415^{th}$, and $430^{th}$ day, showing the appearance of instability.}
\end{figure}

\
\section{Conclusions and Discussion} \label{sec:floats}

In the closed ocean basin, the cosine wind we choose generates a single gyre initially. Small planetary $\beta$ favors the acceleration of the ocean circulation and the broadening of the western boundary layer; the decrease of eddy viscosity $A_h$, which means less dissipation, also speeds up the ocean flow while shrinks the western boundary layer. The variation of wind stress $\tau$, which directly drives the circulation, influences the strength of the ocean velocities but not the thickness of the western boundary. In the channel ocean, the same wind stress produces zonal currents. Planetary $\beta$ doesn't significantly influence the ocean currents in the channel ocean, while the eddy viscosity and wind stress affect the zonal velocity and the sea surface height. When the specified wind satisfies the necessary condition of barotropic instability, instability occurs in the channel ocean. In the barotropic channel ocean, the zonal currents can develop turbulence only when it satisfies the necessary condition of barotropic instability. While in a closed basin ocean, the existence of east and west boundaries makes ocean circulations turn into unsteady states easily as long as the current velocity is large enough.

The real ocean is stratified, has complicated land configurations and is forced by wind stress and heat flux with rich structures at a range of spatiotemporal scales. The idealized model we employ here is much simpler than realistic three-dimensional fully coupled global models and the real ocean. Conclusions here may not exactly hold in a realistic model. However, the sensitivity of ocean circulations to various parameters, revealed from our simple model, can shed light on further investigations and understandings about ocean circulations, planetary climates and habitability. Changes of ocean circulations might greatly influence the planetary climate through transporting heat, carbon and nutrients. Future work using fully coupled ocean-atmosphere models are needed to understand the coupling between the four different components of the climate system and the net effect of ocean circulations on planetary climates.

\acknowledgments

\textbf{Acknowledgments:} J.Y. acknowledges supports from the National Science Foundation of China (NSFC) under grants 41675071, 41606060, 41761144072 and 4171101348.

\
\
\

 \bibliographystyle{aasjournal}
%
\bibliography{paper.bbl}

\begin{thebibliography}{}
\expandafter\ifx\csname natexlab\endcsname\relax\def\natexlab#1{#1}\fi
\providecommand{\url}[1]{\href{#1}{#1}}

\bibitem[{Beaulieu {et~al.}(2006)Beaulieu, Bennett, Fouqu{\'e}, Williams,
  Dominik, J{\o}rgensen, Kubas, Cassan, Coutures, Greenhill,
  {et~al.}}]{beaulieu2006discovery}
Beaulieu, J.-P., Bennett, D.~P., Fouqu{\'e}, P., {et~al.} 2006, Nature, 439,
  437

\bibitem[{Boccaletti {et~al.}(2005)Boccaletti, Ferrari, Adcroft, Ferreira, \&
  Marshall}]{boccaletti2005vertical}
Boccaletti, G., Ferrari, R., Adcroft, A., Ferreira, D., \& Marshall, J. 2005,
  Geophysical Research Letters, 32

\bibitem[{Carton {et~al.}(2018)Carton, Chepurin, \& Chen}]{carton2018updated}
Carton, J., Chepurin, G., \& Chen, L. 2018, An updated reanalysis of ocean
  climate using the Simple Ocean Data Assimilation version 3 (SODA3),
  manuscript in preparation.
\newblock \url{http://www.atmos.umd.edu/~ocean/}

\bibitem[{Charney(1955)}]{charney1955gulf}
Charney, J.~G. 1955, Proceedings of the National Academy of Sciences, 41, 731

\bibitem[{Cullum {et~al.}(2016)Cullum, Stevens, \&
  Joshi}]{cullum2016importance}
Cullum, J., Stevens, D.~P., \& Joshi, M.~M. 2016, Proceedings of the National
  Academy of Sciences, 113, 4278

\bibitem[{Del~Genio {et~al.}(2017)Del~Genio, Way, Amundsen, Aleinov, Kelley,
  Kiang, \& Clune}]{del2017habitable}
Del~Genio, A.~D., Way, M.~J., Amundsen, D.~S., {et~al.} 2017, arXiv preprint
  arXiv:1709.02051

\bibitem[{Ferreira \& Marshall(2006)}]{ferreira2006formulation}
Ferreira, D., \& Marshall, J. 2006, Ocean Modelling, 13, 86

\bibitem[{Gill(1982)}]{gill1982atmosphere}
Gill, A.~E. 1982, International Geophysics Series

\bibitem[{Hecht(2016)}]{hecht2016truth}
Hecht, J. 2016, Nature, 530, 272

\bibitem[{Hu \& Cui(1991)}]{dunxin1991western}
Hu, D., \& Cui, M. 1991, Chinese Journal of Oceanology and Limnology, 9, 1

\bibitem[{Hu {et~al.}(2015)Hu, Wu, Cai, Gupta, Ganachaud, Qiu, Gordon, Lin,
  Chen, Hu, {et~al.}}]{hu2015pacific}
Hu, D., Wu, L., Cai, W., {et~al.} 2015, Nature, 522, 299

\bibitem[{Hu \& Yang(2014)}]{hu2014role}
Hu, Y., \& Yang, J. 2014, Proceedings of the National Academy of Sciences, 111,
  629

\bibitem[{Huang(2010)}]{huang2010ocean}
Huang, R.~X. 2010, Ocean circulation: wind-driven and thermohaline processes
  (Cambridge University Press)

\bibitem[{L{\'e}ger {et~al.}(2004)L{\'e}ger, Selsis, Sotin, Guillot, Despois,
  Mawet, Ollivier, Lab{\`e}que, Valette, Brachet, {et~al.}}]{leger2004new}
L{\'e}ger, A., Selsis, F., Sotin, C., {et~al.} 2004, Icarus, 169, 499

\bibitem[{Lund {et~al.}(2006)Lund, Lynch-Stieglitz, \& Curry}]{lund2006gulf}
Lund, D.~C., Lynch-Stieglitz, J., \& Curry, W.~B. 2006, Nature, 444, 601

\bibitem[{Marshall {et~al.}(2007)Marshall, Ferreira, Campin, \&
  Enderton}]{marshall2007mean}
Marshall, J., Ferreira, D., Campin, J.-M., \& Enderton, D. 2007, Journal of the
  Atmospheric Sciences, 64, 4270

\bibitem[{Mcwilliams {et~al.}(1978)Mcwilliams, Holland, \&
  Chow}]{mcwilliams1978description}
Mcwilliams, J.~C., Holland, W.~R., \& Chow, J.~H. 1978, Dynamics of Atmospheres
  and Oceans, 2, 213

\bibitem[{Munk(1950)}]{munk1950wind}
Munk, W.~H. 1950, Journal of Meteorology, 7, 80

\bibitem[{Nadeau \& Ferrari(2015)}]{nadeau2015role}
Nadeau, L.-P., \& Ferrari, R. 2015, Journal of Physical Oceanography, 45, 1491

\bibitem[{Orsi {et~al.}(1995)Orsi, Whitworth, \& Nowlin}]{orsi1995meridional}
Orsi, A.~H., Whitworth, T., \& Nowlin, W.~D. 1995, Deep Sea Research Part I:
  Oceanographic Research Papers, 42, 641

\bibitem[{O’Neill {et~al.}(2015)O’Neill, Emanuel, \& Flierl}]{o2015polar}
O’Neill, M.~E., Emanuel, K.~A., \& Flierl, G.~R. 2015, Nature Geoscience, 8,
  523

\bibitem[{O’Neill {et~al.}(2016)O’Neill, Emanuel, \& Flierl}]{o2016weak}
---. 2016, Journal of the Atmospheric Sciences, 73, 1841

\bibitem[{Pickard \& Emery(2016)}]{pickard2016descriptive}
Pickard, G.~L., \& Emery, W.~J. 2016, Descriptive physical oceanography: an
  introduction (Elsevier)

\bibitem[{Rayleigh(1879)}]{rayleigh1879stability}
Rayleigh, L. 1879, Proceedings of the London Mathematical Society, 1, 57

\bibitem[{Stommel(1948)}]{stommel1948westward}
Stommel, H. 1948, Eos, Transactions American Geophysical Union, 29, 202

\bibitem[{Talley(2011)}]{talley2011descriptive}
Talley, L.~D. 2011, Descriptive physical oceanography: an introduction
  (Academic press)

\bibitem[{Trenberth \& Caron(2001)}]{trenberth2001estimates}
Trenberth, K.~E., \& Caron, J.~M. 2001, Journal of Climate, 14, 3433

\bibitem[{Vallis(2017)}]{vallis2017atmospheric}
Vallis, G.~K. 2017, Atmospheric and oceanic fluid dynamics (Cambridge
  University Press)

\bibitem[{Whitworth~III \& Peterson(1985)}]{whitworth1985volume}
Whitworth~III, T., \& Peterson, R. 1985, Journal of Physical Oceanography, 15,
  810

\bibitem[{Winton(2003)}]{winton2003climatic}
Winton, M. 2003, Journal of Climate, 16, 2875

\bibitem[{Wunsch(2011)}]{wunsch2011decadal}
Wunsch, C. 2011, Journal of Marine Research, 69, 417

\end{thebibliography}

\end{document}